\documentclass{article}
\usepackage{spconf,amsmath,graphicx,hyperref}
\usepackage{multicol}   
\usepackage{caption}    
\usepackage{tabularx}
\usepackage{amssymb}
\usepackage{booktabs}
\usepackage{multirow}
\usepackage{fancyhdr}  

\title{PoolingVQ: A VQVAE Variant for Reducing Audio Redundancy and Boosting Multi-Modal Fusion in Music Emotion Analysis}
\name{Dinghao Zou$^{\star \dagger}$ \qquad Yicheng Gong$^{\star} $ \qquad Xiaokang Li$^{\dagger} $ \qquad Xin Cao$^{\dagger} $ \qquad Sunbowen Lee$^{\dagger} $
}
\address{$^{\star \dagger}$ Wuhan University of Science and
Technology, Wuhan, China }
\begin{document}
%
\maketitle
\begin{abstract}
Multimodal music emotion analysis leverages both audio and MIDI modalities to enhance performance. While mainstream approaches focus on complex feature extraction networks, we propose that shortening the length of audio sequence features to mitigate redundancy, especially in contrast to MIDI's compact representation, may effectively boost task performance.
To achieve this, we developed \textbf{PoolingVQ} by combining Vector Quantized Variational Autoencoder (\textbf{VQVAE}) with \textbf{spatial pooling}, which directly compresses audio feature sequences through codebook-guided local aggregation to reduce redundancy, then devised a two-stage co-attention approach to fuse audio and MIDI information. Experimental results on the public datasets EMOPIA and VGMIDI demonstrate that our multimodal framework achieves state-of-the-art performance, with PoolingVQ yielding effective improvement. Our proposed method's code is available at Anonymous GitHub\footnote{\url{https://anonymous.4open.science/r/poolingvq}}.
\end{abstract}
\section{Introduction}
\label{sec:intro}
Music Emotion Analysis (MEA) models music-emotion links, enabling machines to perceive/quantify affect and supporting applications like recommendation and therapy via accurate emotional interpretation.
Constrained by scarce datasets, current MEA focuses on utterance-level tasks \cite{0}: classification (Russell’s quadrant model, arousal/valence) and regression (predicting scores for these dimensions)—adapting to data limits while decoding emotions.

Early MEA methods used rule-based approaches with musicological features (mode, tempo, harmony \cite{1, 2, 3}) —interpretable but limited by handcrafted features and expert dependence, restricting expressiveness and generalization. Deep learning enabled CNN/RNN-based models to capture abstract audio/lyric representations \cite{4, 5, 6}, yet most focused on single modalities, missing multi-signal emotional cues.
Advances in deep learning brought multimodal architectures integrating audio, MIDI and lyrics \cite{7, 8, 9}, while pre-training technology facilitated the emergence of strong representation models: MusicBERT \cite{10} (transformer-based, symbolic music) captures structure and harmony; MERT \cite{11} (diverse audio pre-training) provides
rich acoustic representations. \textbf{However}, most existing methods still rely primarily on feature extraction techniques, with little consideration given to the redundancy in audio features \cite{liu2023simple,liu2024} and the cross-modal sequence length discrepancy between audio and MIDI features \cite{zou1,zou2}.

\begin{figure*}[t]
    \centering
    \includegraphics[width=\textwidth]{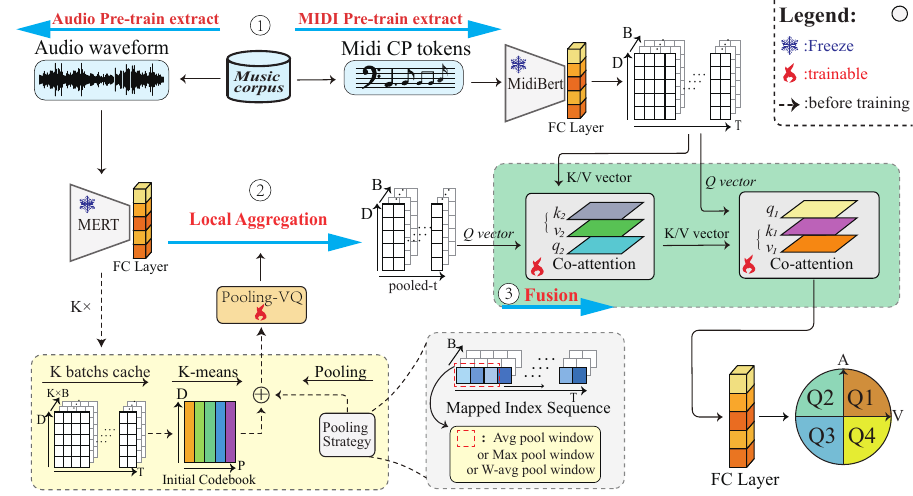} 
     \caption{Architectures of our proposed model, which consists of three modules: pre-train model extractor module, music audio representation Local Aggregation module and multi-modal fusion module.}  
    \label{fig:model_architecture}
\end{figure*}

In this case, we focus on sequence length adjustment through local aggregation to reduce audio redundancy and cross-modal length discrepancies. \textbf{We propose a novel VQVAE variant: PoolingVQ}, \textbf{which is inspired by VQVAE’s local mapping capability in quantization \cite{14} and the aggregation effects of spatial pooling methods \cite{pool}}. Our framework is shown in Figure \ref{fig:model_architecture}.

\begin{figure}[http]  
    \centering  
    \includegraphics[width=\linewidth]{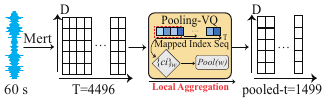}  
    \caption{Changes of a single sample via PoolingVQ.}
    \label{fig:audio-length}
\end{figure}

Specifically, the PoolingVQ framework consists of the following three key steps: 1) initializing the VQVAE codebook via K-means clustering on K batches of MERT-extracted audio features. 2) generating index sequences by mapping local features through the codebook during training. 3) applying a sliding window to these sequences, with index variations guiding optimal local feature pooling. \textbf{As shown in Figure \ref{fig:audio-length}}, the length of audio features (max 60s duration) is shortened to a third of the original; similarly, the length gap relative to MIDI features (length fixed at 512) is reduced to a third of its original magnitude.
Subsequently, compressed audio is then fused with MIDI-BERT
features via a two-stage co-attention fusion module, finally a classifier for emotion prediction.

To evaluate the effectiveness of PoolingVQ, we conducted test and abltion experiments on two diverse datasets (EMOPIA: pure piano; VGMIDI: game soundtracks). Results show PoolingVQ reduces audio features length while boosting performance. Compared to blind aggregation with only average/max pooling, 
the codebook-guided local mapping significantly improves the F1-score, with EMOPIA achieving around 89.5\% and VGMIDI reaching up to approximately 55\%, and there exists a stable performance peak interval.

In summary, our contributions are threefold: 1) Proposing a novel VQVAE variant via codebook-guided local aggregation for audio redundancy reduction. 2) Integrating pre-trained musical representations with multi-modal modeling. 3) Validating that local audio feature aggregation enhances multi-modal music emotion classification.

\section{Method}
\label{sec:format}



As shown in Figure 1, our model comprises three modules: pre-trained extractor, audio Local Aggregation, and multi-modal fusion.
1) The pre-trained extractor uses frozen MERT and MIDIBERT to extract audio and MIDI features, respectively. 2) The PoolingVQ-based Local Aggregation processes audio branch: MERT features are codebook-mapped into local index sequences, then these MERT features (guided by the variations in the local index sequences) undergo pooling operations (max, average, or weighted average).
3) Pooled audio and MIDI features enter a fusion module with two-stage co-attentions. Fused features are fed to a classifier for emotion prediction.

\subsection{Pre-trained Model Extractor Module}
This module employs frozen pre-trained models MERT-95M and MIDI-BERT to extract feature representations for audio and MIDI modalities, respectively. For the audio branch, the input audio data is processed according to the specifications of MERT-95M, with the sampling rate converted to 22400Hz. For the MIDI branch, we retained MIDI-BERT's pre-trained parameters. To align with the MIDI branch (the maximum sequence length of 512) and avoid random audio truncation, we analyzed audio duration distributions of effective data lengths from EMOPIA \cite{12} and VGMIDI \cite{13}, then set maximum audio durations to 60s (EMOPIA) and 130s (VGMIDI) for cross-modal temporal consistency.

\subsection{Local Aggregation Module}
The pooling compression module is centered on PoolingVQ, combining the VQVAE codebook and pooling strategies. It accepts audio representations extracted from the frozen pre-trained model MERT-95M.

In this work, VQVAE models the local feature distributions of $\mathbf{X} = \{ x_i \in \mathbb{R}^{D}, \, i = 1, \ldots, T \}$, utilizing index sequences to capture local variations and determine corresponding pooling strategies. The framework involves two stages: pre-training, where K batches of data are cached for k-means clustering to generate p cluster centers as VQVAE's initial codebook P; and formal training, employing classical VQVAE mapping with precomputed similarity (higher similarity guiding better mapping):
\begin{equation}
\cos(x_i, P_j) = \frac{x_i \cdot P_j}{\|x_i\| \|P_j\|},
\end{equation}
\begin{equation}
c_i = \arg\max \left( \cos(x_i, P_1),...,\cos(x_i, P_p)\right),
\end{equation}
where cos($x_i$, $P_j$) is the cosine similarity between local feature $x_i$ and the $j$-th codebook vector $P_j$, and $c_i$ denotes the index of the codebook vector assigned $x_i$ (i.e., the index corresponding to the maximum cosine similarity).

By conducting local mapping on X, we obtain mapped index sequence 
$\mathbf{C} = \{ c_i \in \mathbb{N}, \, i = 1, \ldots, T \}$. Then a piece-wise pooling strategy $Pool(w)$ is defined for a sliding window $w$ (size 5, stride 3) based on the length of mapped index sequence $\{c_i\}_w$, which  
is the index set in $w$. 
\begin{equation}
Pool(w) = 
\begin{cases} 
AvgP & |\{c_i\}_w| = 1, \\
WtAvgP & 1 < |\{c_i\}_w| \leq 3, \\
MaxP & |\{c_i\}_w| > 3,
\end{cases}
\label{eq:pooling_strategy}
\end{equation}
where $AvgP$, $WtAvgP$, $MaxP$ denote Average, Weighted Average, Max Pooling respectively. The sequence changes are shown in Figure 2.

To enable more accurate modeling of local features, the traditional VQVAE \cite{14} quantization loss fine-tunes cluster center positions.
\begin{equation}
\mathcal{L}_1 = \|sg(X) - P_c\|_2 + \|X - sg(P_c)\|_2 ,
\end{equation}
where \( X \) is the original audio input, \( P_c \) is the codebook-mapped feature vector for the mapped index sequence \( C \), and \( sg(\cdot) \) is the stop-gradient operation blocking gradient flow through its parameters in backpropagation.

\subsection{Multi-modal Fusion Module}
Modal fusion matters as heterogeneous modalities (e.g., audio, text) offer complementary info, bridging single-modal gaps. Unlike simple uniform methods (concatenation, element-wise addition), co-attention dynamically models inter-modal links via adaptive weights for relevant pairs, cutting noise by focusing on key correlations through query-key scoring and value aggregation. The co-attention calculation formula is:
\begin{equation}
Attn(Q_\alpha, K_\beta, V_\beta) = softmax\left( \frac{Q_\alpha K_\beta^T}{\sqrt{d}} \right) V_\beta ,
\end{equation}
where $\alpha$, $\beta$ denote different modalities.
In this paper, Building on \cite{15,16}, we propose a two-stage co-attention fusion module: first, compressed audio (as queries) fuses with MIDI features; then the process reverses. This bidirectional interaction yields richer fused features.
\begin{equation}
\begin{gathered}
F_1 = Attn(Q_a, K_m, V_m),\\
F_{final} = Attn(Q_m, K_1, V_1),
\end{gathered}
\end{equation}
where $F_1$ is the initial fused feature from the first-stage co-attention (audio query with MIDI key/value), and $F_{final}$ is the final fused feature from the second-stage co-attention (MIDI query with $F_1$ key/value).

The final fused feature $F_{final}$ is fed into a fully connected classification head, outputting raw prediction values $z_j$ for each class $j$. For the four-class task, softmax is applied to $z_j$ to get predicted probabilities $\hat{y}_j$ ($j=1,\dots,4$). Specifically, cross-entropy loss measures the discrepancy between $\hat{y}_j$ and ground-truth labels $y_j$, defined as:

\begin{equation}
\mathcal{L}_{CE} = -\sum_{j = 1}^{4} y_j \log(\hat{y}_j).
\end{equation}
Therefore, the final total loss of our model is:
\begin{equation}
\mathcal{L}_{total} = \mathcal{L}_{CE} + \mathcal{L}_1 .
\end{equation}

\section{EXPERIMENTS}
\label{sec:pagestyle}

\subsection{Experimental Setup}
We trained and validated our model using the EMOPIA and VGMIDI datasets. Both datasets are used for a four-classification task. Among them, EMOPIA consists of pure piano music with 1087 samples, and VGMIDI contains 200 samples of game soundtracks. To ensure comparability with previous methods, we followed the official division files of the original data: EMOPIA was divided in a 7:2:1 ratio, and VGMIDI in a 6:2:2 ratio. As noted in Section 2.1, the maximum audio duration is 60s for EMOPIA and 130s for VGMIDI, with the maximum MIDI sequence length at 512.

The K-batch buffer caches half of the training data.
Training was conducted using the Adam optimizer with parameters set as $\beta_1$=0.9, $\beta_2$=0.99, $\varepsilon$=1e-6. Specifically, the classifier adopted a learning rate of 1e-4 and weight decay of 5e-5, while the quantizer used a learning rate of 5e-5 and weight decay of 1e-5. Additionally, the ReduceLROnPlateau scheduling strategy was incorporated, featuring a decay factor of 0.7, a patience of 2 epochs, and a threshold of 1e-4.

\subsection{Results and Analysis}
\begin{table}[http] 
    \raggedright
    \normalsize
    \captionsetup{
        justification=raggedright, 
        singlelinecheck=false,
        font=normalsize
    }
    \caption{Performance Comparison of Different Methods on EMOPIA and VGMIDI Datasets.}
    \label{tab:performance_comparison}
    \heavyrulewidth=0.1em 
    \lightrulewidth=0.05em  
    \begin{tabular}{lcccc}
        \toprule 
        \textbf{Method} & \multicolumn{2}{c}{\textbf{EMOPIA}} & \multicolumn{2}{c}{\textbf{VGMIDI}} \\
        \cmidrule(lr){2-3} \cmidrule(lr){4-5} 
        & \textit{Acc} & \textit{F1} & \textit{Acc} & \textit{F1} \\
        \midrule 
        SVM \cite{2} & 0.477 & 0.476 & 0.451 & 0.377 \\
        LSTM-Ann \cite{12} & 0.647 & 0.563 & 0.417 & 0.260 \\
        MT-MIDIGPT \cite{8} & 0.625 & 0.611 & \underline{0.558} & 0.509 \\
        MT-MIDIBERT \cite{8} & 0.676 & 0.664 & 0.498 & 0.453 \\
        BFAM \cite{7} & \underline{0.822} & \underline{0.770} & \textbf{0.625} & \underline{0.547} \\
        \midrule 
        Our proposed model & \textbf{0.8953} & \textbf{0.8954} & 0.550 & \textbf{0.5501} \\
        \bottomrule 
    \end{tabular}
\end{table}

For a comprehensive comparison with existing approaches, we evaluated representative methods spanning traditional, single-modal neural network, pre-trained model, and multi-modal methods, with results summarized in Table \ref{tab:performance_comparison}. Our model achieves an F1-score of 89.54\% on the EMOPIA dataset and 55.01\% on VGMIDI, collectively delivering state-of-the-art  performance.
Among the comparative methods, the traditional approach SVM and the MIDI-based single-modal method LSTM-Ann exhibit limited efficacy, while pre-trained models demonstrate measurable performance gains. As a multi-modal approach, BFAM stands out as the one with the most significant progress in overall performance. By integrating pre-trained musical representations with multi-modal modeling (to capture emotional features in both acoustic and symbolic domains), our model demonstrates strong competitiveness. This validates that combining pre-training with multi-modal fusion effectively enhances music emotion recognition performance, offering a more advantageous technical pathway for the field.

\subsection{Ablation Study}

\begin{figure}[t]  
    \centering  
    \includegraphics[width=\linewidth]{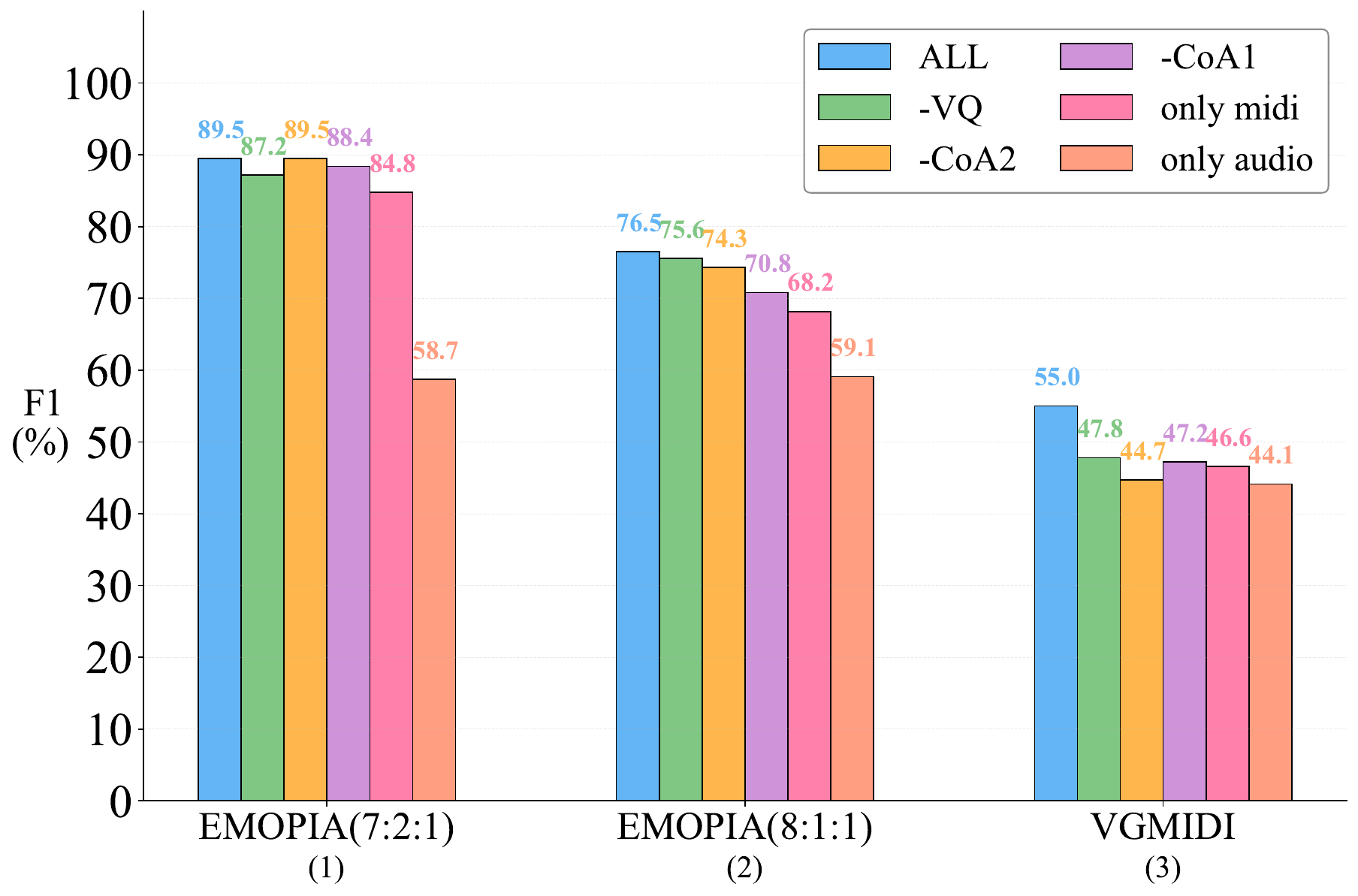}  
    \caption{Bar chart of ablation experiment F1-scores. (a) Experiments (1) and (2) test EMOPIA's stability under different splits. (b) Experiments (1) and (3) test stability across EMOPIA and VGMIDI.}  
    \label{fig:ablation_study1}  
\end{figure}

1) We evaluated three key components: Local Aggregation module \textbf{VQ}, co-attention of the first stage \textbf{CoA1}, co-attention of the second stage \textbf{CoA2}. And the performance of unimodal setups were compared, with results in Figure \ref{fig:ablation_study1}.

Ablation (\textit{a}) on EMOPIA was conducted to test the impact of different splitting methods: the “ALL” configuration consistently achieves the highest F1-score, and other configurations show relatively consistent performance patterns, which indicates consistent impacts of components across splits. Ablation (\textit{b}) was conducted on EMOPIA and the more challenging VGMIDI to test the impact of different datasets. : “ALL” remains optimal, gaining significantly over single-modal setups (only midi, only audio), validating consistent component effectiveness across data sets. \textbf{On the whole}, “ALL” outperforms `-vq' to different extents, showing that our PoolingVQ reduces audio redundancy while enhancing performance.

\begin{figure}[htbp]  
    \centering  
    \includegraphics[width=\linewidth]{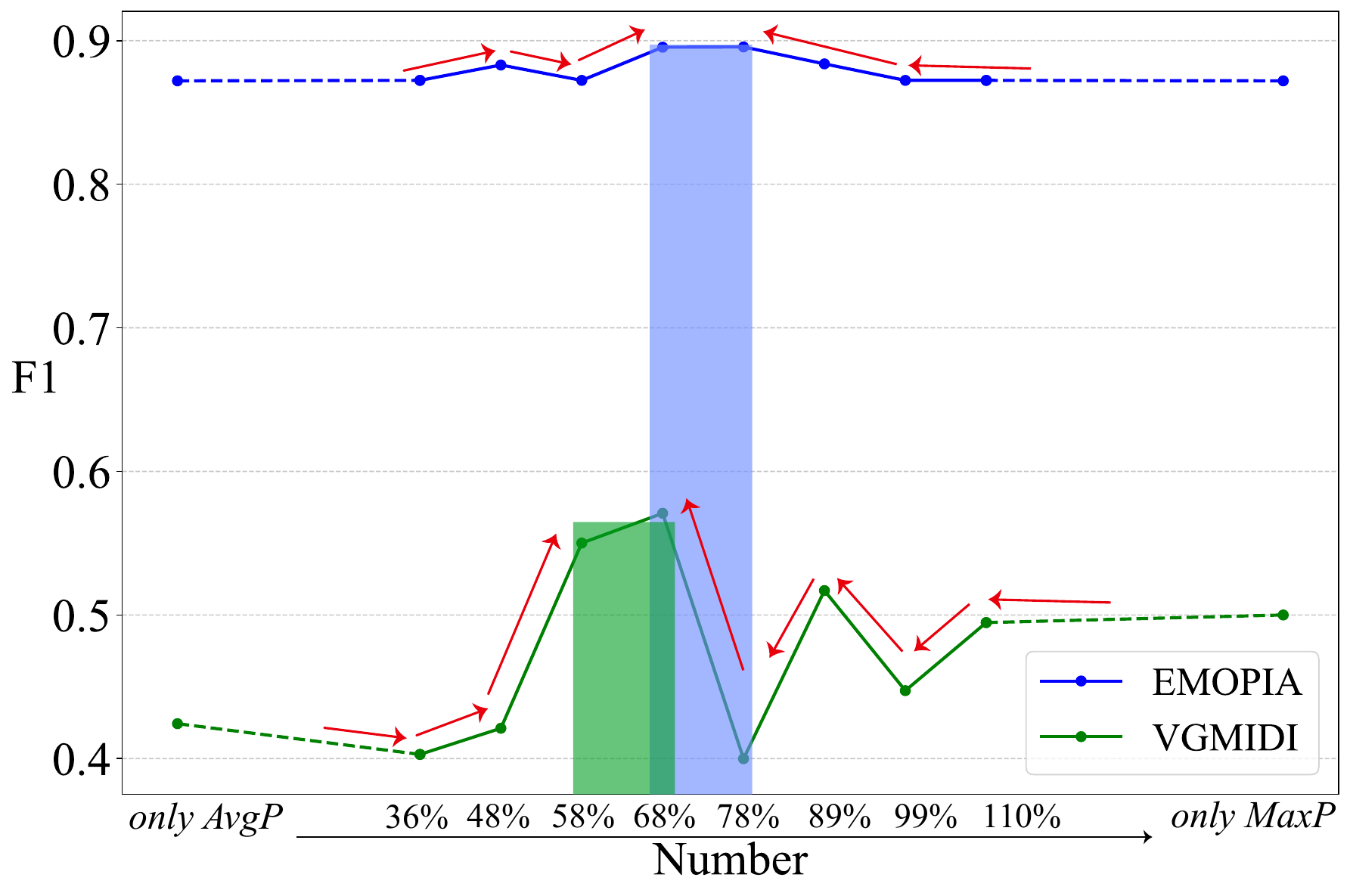}  
    \caption{Impact of Codebook Number (P): The X-axis represents the percentage of the number of codebooks relative to the length of the MERT audio feature sequence.}  
    \label{fig:ablation_study2}  
\end{figure}

2) Impact of initial codebook number (parameter p): we explored p's influence by taking various values on the EMOPIA and VGMIDI datasets, with results in Figure \ref{fig:ablation_study2}.

As the X-axis endpoints, ``Only AvgP'' and ``Only MaxP'' represent extreme states with extremely small and large codebook number, respectively. On both EMOPIA and VGMIDI datasets, performance gradually rises as codebook number approach intermediate values. \textbf{This indicates that}, compared to a single and arbitrary aggregation approach, the mapped sequences of VQVAE codebooks can guide the selection of optimal aggregation strategies. \textbf{Additionally}, considering overall the peak intervals of the two datasets, although the optimal percentage ranges for datasets from different domains vary, they still show slight overlap. This suggests that the method may generally have a consistent effective range of 60\%-80\% across these datasets.

\section{CONCLUSION and future work}
\label{sec:typestyle}
In this study, we propose PoolingVQ—a novel VQVAE variant that integrates a pooling strategy for local audio feature aggregation. Designed to balance audio redundancy reduction and feature preservation, PoolingVQ enhances subsequent two-stage multi-modal fusion, with efficacy demonstrated across two distinct-domain datasets.
Some future work can involve exploring local aggregation with codebooks optimized for emotional info, aligning better with music’s emotional transition continuity—beyond just local variations.

\section{ACKNOWLEDGEMENT}
\label{sec:majhead}

This work was supported by the National Natural Science Foundation of China (NO.12171378) and High
Performance Computing Center of Wuhan University of Science
and Technology. We thank the reviewers for their contributions.





\AtBeginEnvironment{thebibliography}{
  \baselineskip=11pt 
}

\bibliographystyle{IEEEbib}
\bibliography{strings,refs}

\begin{thebibliography}{10}

\bibitem{0}
J.~Kang and D.~Herremans,
\newblock ``Are we there yet? a brief survey of music emotion prediction datasets, models and outstanding challenges,''
\newblock {\em IEEE Transactions on Affective Computing}, pp. 1--16, 2025.

\bibitem{1}
S.~Beveridge and D.~Knox,
\newblock ``Popular music and the role of vocal melody in perceived emotion,''
\newblock {\em Psychology of Music}, vol. 46, no. 3, pp. 411--423, 2018.

\bibitem{2}
Y.~Lin, X.~Chen, and D.~Yang,
\newblock ``Exploration of music emotion recognition based on midi,''
\newblock {\em Proceedings of the International Society for Music Information Retrieval Conference}, pp. 221--226, 2013.

\bibitem{3}
K.W. Cheuk, Y.-J. Luo, B.~Balamurali, G.~Roig, and D.~Herremans,
\newblock ``Regression-based music emotion prediction using triplet neural networks,''
\newblock in {\em International Joint Conference on Neural Networks}, 2020, pp. 1--7.

\bibitem{4}
Y.~Dong, X.~Yang, X.~Zhao, and J.~Li,
\newblock ``Bidirectional convolutional recurrent sparse network (bcrsn): an efficient model for music emotion recognition,''
\newblock {\em IEEE Transactions on Multimedia}, vol. 21, no. 12, pp. 3150--3163, 2019.

\bibitem{5}
S.~Rajesh and N.J. Nalini,
\newblock ``Musical instrument emotion recognition using deep recurrent neural network,''
\newblock {\em Procedia Computer Science}, vol. 167, pp. 16--25, 2020.

\bibitem{6}
S.~Chaki, P.~Doshi, P.~Patnaik, and S.~Bhattacharya,
\newblock ``Attentive rnns for continuous-time emotion prediction in music clips,''
\newblock in {\em AffCon@ AAAI}, 2020, pp. 36--46.

\bibitem{7}
Y.~Xiao, H.~Ruan, X.~Zhao, P.~Jin, L.~Tian, Z.~Wei, X.~Cai, Y.~Wang, and L.~Liu,
\newblock ``An efficient bi-modal fusion framework for music emotion recognition,''
\newblock {\em IEEE Transactions on Affective Computing}, pp. 1--17, 2024.

\bibitem{8}
J.~Qiu, C.L. Chen, and T.~Zhang,
\newblock ``A novel multi-task learning method for symbolic music emotion recognition,''
\newblock {\em arXiv preprint arXiv:2201.05782}, 2022.

\bibitem{9}
J.~Zhao, G.~Ru, Y.~Yu, Y.~Wu, D.~Li, and W.~Li,
\newblock ``Multimodal music emotion recognition with hierarchical cross-modal attention network,''
\newblock in {\em Proc. IEEE Int. Conf. Multimedia Expo}, Taipei, Taiwan, 2022, IEEE, pp. 1--6.

\bibitem{10}
H.Y. Zhu, Y.~Niu, D.~Fu, and H.~Wang,
\newblock ``Musicbert: A self-supervised learning of music representation,''
\newblock in {\em ACM International Conference on Multimedia}, 2021, pp. 3955--3963.

\bibitem{11}
Y.~Li, R.B. Yuan, G.~Zhang, Y.H. Ma, X.R. Chen, H.Z. Yin, C.H. Xiao, C.H. Lin, A.~Ragni, E.~Benetos, et~al.,
\newblock ``Mert: Acoustic music understanding model with large-scale self-supervised training,''
\newblock in {\em The Twelfth International Conference on Learning Representations}, 2024.

\bibitem{liu2023simple}
X.B. Liu, H.H. Liu, Q.Q. Kong, X.H. Mei, M.D. Plumbley, and W.W. Wang,
\newblock ``Simple pooling front-ends for efficient audio classification,''
\newblock in {\em ICASSP 2023-2023 IEEE International Conference on Acoustics, Speech and Signal Processing (ICASSP)}. IEEE, 2023, pp. 1--5.

\bibitem{liu2024}
J.~Feng, M.H. Erol, J.S. Chung, and A.~Senocak,
\newblock ``Elasticast: An audio spectrogram transformer for all length and resolutions,''
\newblock in {\em Proceedings of the Annual Conference of the International Speech Communication Association, INTERSPEECH}, 2024, pp. 4743--4747.

\bibitem{zou1}
C.L. Li, H.Y. Xu, J.F. Tian, W.~Wang, M.~Yan, B.~Bi, J.B. Ye, H.~Chen, G.H. Xu, Z.~Cao, et~al.,
\newblock ``mplug: Effective and efficient vision-language learning by cross-modal skip-connections,''
\newblock in {\em Proceedings of the 2022 Conference on Empirical Methods in Natural Language Processing}, dec 2022, pp. 7241--7259.

\bibitem{zou2}
Z.~Fu, F.~Liu, Q.~Xu, X.~Fu, and J.~Qi,
\newblock ``Lmr-cbt: Learning modality-fused representations with cb-transformer for multimodal emotion recognition from unaligned multimodal sequences,''
\newblock {\em Frontiers of Computer Science}, vol. 18, no. 4, pp. 184314, 2024.

\bibitem{14}
A.~van~den Oord, O.~Vinyals, and K.~Kavukcuoglu,
\newblock ``Neural discrete representation learning,''
\newblock {\em Advances in Neural Information Processing Systems}, vol. 30, 2017.

\bibitem{pool}
A.~Stergiou and R.~Poppe,
\newblock ``Adapool: Exponential adaptive pooling for information-retaining downsampling,''
\newblock {\em IEEE Transactions on Image Processing}, vol. 32, pp. 251--266, 2022.

\bibitem{12}
H.T. Hung, J.~Ching, S.H. Doh, N.~Kim, J.~Nam, and Y.H. Yang,
\newblock ``Emopia: A multi-modal pop piano dataset for emotion recognition and emotion-based music generation,''
\newblock in {\em International Society for Music Information Retrieval Conference, ISMIR 2021}, 2021.

\bibitem{13}
L.N. Ferreira and J.~Whitehead,
\newblock ``Learning to generate music with sentiment,''
\newblock in {\em International Society for Music Information Retrieval Conference}, 2021.

\bibitem{15}
Y.~Wu, P.W. Zhan, Y.J. Zhang, L.M. Wang, and Z.~Xu,
\newblock ``Multimodal fusion with co-attention networks for fake news detection,''
\newblock in {\em Annual Meeting of the Association for Computational Linguistics}, 2021, pp. 2560--2569.

\bibitem{16}
G.H. Ru et~al.,
\newblock ``Improving music genre classification from multi-modal properties of music and genre correlations perspective,''
\newblock in {\em IEEE International Conference on Acoustics, Speech, and Signal Processing}, 2023, pp. 1--5.

\end{thebibliography}


\end{document}